\documentclass[preprintnumbers,amsmath,amssymbm,prd]{revtex4}
\usepackage{epsfig}
\usepackage{graphicx}

\begin{document}

\title{Upper bound on the center-of-mass energy of the collisional Penrose process}
\author{Shahar Hod}
\address{The Ruppin Academic Center, Emeq Hefer 40250, Israel}
\address{ }
\address{The Hadassah Institute, Jerusalem 91010, Israel}
\date{\today}

\begin{abstract}
\ \ \ Following the interesting work of Ba\~nados, Silk, and West
[Phys. Rev. Lett. {\bf 103}, 111102 (2009)], it is repeatedly stated
in the physics literature that the center-of-mass energy, ${\cal
E}_{\text{c.m}}$, of two colliding particles in a maximally rotating
black-hole spacetime can grow unboundedly. For this extreme scenario
to happen, the particles have to collide at the black-hole horizon.
In this paper we show that Thorne's famous hoop conjecture precludes
this extreme scenario from occurring in realistic black-hole
spacetimes. In particular, it is shown that a new (and larger)
horizon is formed {\it before} the infalling particles reach the
horizon of the original black hole. As a consequence, the
center-of-mass energy of the collisional Penrose process is {\it
bounded} from above by the simple scaling relation ${\cal
E}^{\text{max}}_{\text{c.m}}/2\mu\propto(M/\mu)^{1/4}$, where $M$
and $\mu$ are respectively the mass of the central black hole and
the proper mass of the colliding particles.
\end{abstract}
\bigskip
\maketitle


\section{Introduction}

The collisional Penrose process, a collision between two particles
which takes place within the ergosphere of a spinning Kerr black
hole, has attracted the attention of physicists ever since the
pioneering work of Piran, Shaham, and Katz more than four decades
ago \cite{PSK,PS}. Interestingly, it was shown \cite{PSK,PS} that
the collision of two particles in the black-hole spacetime can
produce two new particles, one of which may escape to infinity while
carrying with it some of the black-hole rotational energy
\cite{Noteneg,Pen,NotePen}.

The collisional Penrose process has recently gained renewed interest
when Ba\~nados, Silk, and West \cite{Ban1} have revealed the
interesting fact that the center-of-mass energy of the two colliding
particles may {\it diverge} if the collision takes place at the
horizon of a maximally rotating (extremal) black hole. In
particular, it was suggested \cite{Ban1} that these extremely
energetic near-horizon collisions may provide a unique probe of the
elusive Planck-scale physics \cite{Notenf,Ban2}.

The intriguing discovery of \cite{Ban1}, according to which black
holes may serve as extremely energetic particle accelerators, has
sparked an enormous excitement in the physics community. In
particular, following \cite{Ban1} it is repeatedly stated in the
physics literature that the center-of-mass energy, ${\cal
E}_{\text{c.m}}$, of two colliding particles in a maximally rotating
black-hole spacetime can grow unboundedly.

It is worth emphasizing again that, for the center-of-mass energy of
the colliding particles to diverge, the particles have to collide
exactly at the black-hole horizon. The main goal of the present
paper is to reveal the fact that Thorne's famous hoop conjecture
\cite{Thorne} precludes this extreme scenario from occurring in
realistic black-hole spacetimes. In particular, below we shall show
that a new (and larger) horizon is formed {\it before} the infalling
particles reach the horizon of the original black hole. As a
consequence, the center-of-mass energy of the colliding particles in
the black-hole spacetime {\it cannot} grow unboundedly.

\section{The center-of-mass energy of two colliding particles in the black-hole spacetime}

We shall explore the maximally allowed center-of-mass energy
associated with a collision of two particles that start falling from
rest at infinity towards a maximally rotating (extremal
\cite{Noteexm}) Kerr black hole of mass $M$ and angular momentum
$J=M^2$ \cite{Ban1,Ban2,Ban3,Ban4,Noteunit}. The geodesic motions of
the particles in the black-hole spacetime are characterized by
conserved energies
\begin{equation}\label{Eq1}
E_1=E_2=\mu\
\end{equation}
and conserved angular momenta
\begin{equation}\label{Eq2}
L_1=l_1\cdot M\mu\ \ \ ; \ \ \ L_2=l_2\cdot M\mu\  .
\end{equation}
Geodesic trajectories that extend all the way from spatial infinity
down to the black-hole horizon are characterized by angular momenta
in the bounded regime \cite{Bar,Ban1}
\begin{equation}\label{Eq3}
-2(1+\sqrt{1+J/M^2})\leq l \leq 2(1+\sqrt{1-J/M^2})\  .
\end{equation}
In particular, it was shown in \cite{Ban1} that the center-of-mass
energy of the collision may diverge if one of the colliding
particles (say 1) is characterized by the critical angular momentum
(for $J/M^2=1$)
\begin{equation}\label{Eq4}
l_1=2\  .
\end{equation}
The test-particle approximation implies
\begin{equation}\label{Eq5}
{\bar\mu}\equiv {{\mu}\over{M}}\ll1\  .
\end{equation}

As shown in \cite{Ban1,Ban2,Ban3,Ban4}, the center-of-mass energy of
the collision may diverge if the collision takes place at the
horizon of the black hole. In particular, defining
\begin{equation}\label{Eq6}
x_{\text{c}}\equiv {{r_{\text{c}}-M}\over{M}}
\end{equation}
as the dimensionless collision radius of the two particles, one
finds the leading-order divergent behavior \cite{Ban4}
\begin{equation}\label{Eq7}
{\cal E}^2_{\text{c.m}}=\mu^2\times
{{\beta_{\pm}}(2-l_2)\over{x_{\text{c}}}}+O(1)
\end{equation}
for the center-of-mass energy in the near-horizon $x_{\text{c}}\ll1$
region. As shown in \cite{Ban4}, the dimensionless factor
$\beta_{\pm}$ in (\ref{Eq7}) depends on the sign of the radial
momentum of the first (critical \cite{Notecf}) particle ($+$ for an
outgoing particle and $-$ for an ingoing particle), on the value of
the Carter constant $Q_1$ \cite{Car} which characterizes the polar
geodesic motion of that particle in the Kerr black-hole spacetime,
and on the polar angle $\theta$ which characterizes the collision
point of the two particles in the black-hole spacetime \cite{Ban4}.
Specifically, one finds \cite{Ban4}
\begin{equation}\label{Eq8}
\beta_+={{2\big[2+(2-\tilde Q_1)^{1/2}\big]}\over{1+\cos^2\theta}}\
\ \ \ ; \ \ \ \ \beta_-={{2(2+\tilde
Q_1)^{1/2}}\over{1+\cos^2\theta}}\ ,
\end{equation}
where \cite{Ban4,Notemq}
\begin{equation}\label{Eq9}
0\leq\tilde Q_1\equiv {{Q_1}\over{\mu^2}}\leq2\  .
\end{equation}

Interestingly, as pointed out in \cite{Ban1,Ban2,Ban3,Ban4}, the
center-of-mass energy (\ref{Eq7}) of the colliding particles
diverges {\it if} the collision takes place at the horizon
($x_{\text{c}}\to0$) of the extremal black hole.

\section{The hoop conjecture and the upper bound on the center-of-mass energy of the collisional Penrose process}

In the previous section we have seen that the center-of-mass energy
(\ref{Eq7}) of the colliding particles can diverge. As emphasized in
\cite{Ban1,Ban2,Ban3,Ban4}, this intriguing conclusion rests on the
assumption that the particles can collide exactly at the horizon of
the extremal black hole. In the present section we shall show,
however, that Thorne's famous hoop conjecture \cite{Thorne} implies
that, due to the energy carried by the infalling particles, a {\it
new} (and larger) horizon is formed {\it before} the particles reach
the horizon of the original black hole. This fact implies, in
particular, that the optimal collision \cite{Noteop} between the two
particles {\it cannot} take place exactly at the horizon of the
original black hole.

The hoop conjecture, originally formulated by Thorne more than four
decades ago \cite{Thorne}, asserts that a gravitating system of
total mass (energy) $M$ forms a black hole if its circumference
radius $r_{\text{c}}$ is equal to (or smaller than) the
corresponding horizon radius $r_{\text{Sch}}=2M$ of the
Schwarzschild black-hole spacetime \cite{Teuk,Leon,Hodm}.

In the present study we shall use a {\it weaker} version of the hoop
conjecture \cite{Hodm}. In particular, we conjecture that: A
gravitating system of total mass $M$ and total angular momentum $J$
forms a black hole if its circumference radius $r_{\text{c}}$ is
equal to (or smaller than) the corresponding horizon radius
$r_{\text{Kerr}}=M+\sqrt{M^2-(J/M)^2}$ of the Kerr black-hole
spacetime. That is, we conjecture that \cite{Notepa}
\begin{equation}\label{Eq10}
r_{\text{c}}\leq M+\sqrt{M^2-(J/M)^2}\ \  \Longrightarrow \ \
\text{Black-hole horizon exists}\  .
\end{equation}

In the context of the collisional Penrose process that we consider
here, this version of the hoop conjecture implies that a new (and
larger) horizon is formed if the energetic particles that fall
towards the black hole reach the radial coordinate
$r=r_{\text{hoop}}$, where $r_{\text{hoop}}(E_1,E_2,L_1,L_2)$ is
defined by the familiar functional relation of the Kerr black-hole
spacetime [see Eq. (\ref{Eq10})]
\begin{equation}\label{Eq11}
r_{\text{hoop}}=M+E_1+E_2+\sqrt{(M+E_1+E_2)^2-[(J+L_1+L_2)/(M+E_1+E_2)]^2}\
.
\end{equation}
Taking cognizance of Eqs. (\ref{Eq1}), (\ref{Eq2}), (\ref{Eq4}), and
(\ref{Eq11}), one finds
\begin{equation}\label{Eq12}
r_{\text{hoop}}=M+\sqrt{2(2-l_2)M\mu}+2\mu+O(\mu^2/M)
\end{equation}
for the radius of the new horizon. Assuming that the colliding
particles are not engulfed by an horizon, the relation (\ref{Eq12})
implies [see Eq. (\ref{Eq6})]
\begin{equation}\label{Eq13}
x^{\text{min}}_{\text{c}}=\sqrt{2(2-l_2){\bar\mu}}+2{\bar\mu}
\end{equation}
for the minimally allowed value of the dimensionless collision
radius \cite{Notemx}.

Substituting (\ref{Eq13}) into (\ref{Eq7}) one finds that, in the
collisional Penrose process, the center-of-mass energy of the
colliding particles is bounded from above by the relation
\cite{Noterf1}
\begin{equation}\label{Eq14}
{\cal E}^{\text{max}}_{\text{c.m}}=\mu\times
\sqrt{{{(2-l_2)\beta_{\pm}}\over{\sqrt{2(2-l_2){\bar\mu}}+2{\bar\mu}}}}\
\end{equation}
Assuming the strong inequality $\mu\ll (2-l_2)M$ [see Eq.
(\ref{Eq5}) and Eq. (\ref{Eq19}) below], one can approximate
(\ref{Eq14}) by
\begin{equation}\label{Eq15}
{\cal E}^{\text{max}}_{\text{c.m}}=\mu\times
\Big({{2-l_2}\over{2{\bar\mu}}}\Big)^{1/4}\beta^{1/2}_{\pm}\  ,
\end{equation}
which yields the simple expression \cite{Noterf4}
\begin{equation}\label{Eq16}
{\cal E}^{\text{max}}_{\text{c.m}}=2\gamma\cdot M^{1/4}\mu^{3/4}
\end{equation}
for the maximally allowed center-of-mass energy of the colliding
particles in the black-hole spacetime, where
\begin{equation}\label{Eq17}
\gamma\equiv[(2-l_2)\beta^2_{\pm}/32]^{1/4}\  .
\end{equation}

What is the maximally allowed value of the dimensionless pre-factor
$\gamma$ in (\ref{Eq16})? Inspection of Eq. (\ref{Eq8}) reveals that
the factor $\beta_{\pm}$ is maximized if the collision takes place
at the equatorial plane ($\theta=\pi/2$) of the black hole while the
first particle is on an outgoing trajectory with $\tilde Q_1=0$ [see
Eq. (\ref{Eq9})], in which case one finds [see Eq. (\ref{Eq8})]
\cite{Notepn}
\begin{equation}\label{Eq18}
\beta_{\text{max}}=2(2+\sqrt{2})\  .
\end{equation}
In addition, taking cognizance of the fact that geodesic
trajectories which extend all the way from spatial infinity down to
the black-hole horizon are characterized by angular momenta in the
bounded regime (\ref{Eq3}), one finds (for $J/M^2=1$)
\begin{equation}\label{Eq19}
(2-l_2)_{\text{max}}=2(2+\sqrt{2})\  .
\end{equation}
Substituting (\ref{Eq18}) and (\ref{Eq19}) into (\ref{Eq17}), one
finds \cite{Noterf3}
\begin{equation}\label{Eq20}
\gamma_{\text{max}}={{(2+\sqrt{2})^{3/4}}/{\sqrt{2}}}\ .
\end{equation}

\section{Summary}

Following the important work of Ba\~nados, Silk, and West
\cite{Ban1}, it is repeatedly stated in the physics literature that
the center-of-mass energy of two colliding particles in an extremal
(maximally rotating) black-hole spacetime can diverge. For this
extreme scenario to happen, the particles have to collide exactly at
the horizon of the black hole.

In this paper we have shown that Thorne's famous hoop conjecture
\cite{Thorne} [and also its weaker version (\ref{Eq10})] precludes
this infinite-center-of-mass-energy scenario from occurring in
realistic black-hole spacetimes. In particular, the hoop conjecture
implies that, due to the energy carried by the infalling particles,
a new (and larger) horizon is formed {\it before} the particles
reach the horizon of the original black hole. As a consequence, it
was shown that the optimal collision \cite{Noteop} takes place at
[see Eq. (\ref{Eq12})]
$r^{\text{min}}_{\text{c}}=M+\sqrt{2(2-l_2)M\mu}+2\mu>M$
\cite{Noteot,Noterf2,Noterf5,Beretal}, which implies that the
center-of-mass energy of the colliding particles in the black-hole
spacetime is {\it bounded} from above by the simple scaling relation
[see Eqs. (\ref{Eq16}) and (\ref{Eq20})] \cite{Noteweak}
\begin{equation}\label{Eq21}
{{{\cal
E}^{\text{max}}_{\text{c.m}}}\over{2\mu}}=\gamma_{\text{max}}\cdot\Big({{M}\over{\mu}}\Big)^{1/4}\
.
\end{equation}

For the case of two protons colliding in a supermassive Kerr
black-hole spacetime of $10^9$ solar masses, the bound (\ref{Eq21})
implies ${\cal E}^{\text{max}}_{\text{c.m}}\simeq 10^{14}$TeV for
the maximally allowed center-of-mass energy of the collision.
Interestingly, this energy, though being finite, is still many
orders of magnitude larger than the maximally available
center-of-mass energy in the most powerful man-made accelerators
\cite{Acn}.

\bigskip
\noindent {\bf ACKNOWLEDGMENTS}

This research is supported by the Carmel Science Foundation. I would
like to thank Yael Oren, Arbel M. Ongo, Ayelet B. Lata, and Alona B.
Tea for helpful discussions.


\begin{thebibliography}{99}
\bibitem{Noteneg} Note that the second particle has a negative energy as
measured by asymptotic observers, and it must therefore plunge into
the black hole.
\bibitem{Notenf} It is important to note that subsequent investigations \cite{Ban2} of
the collisional Penrose process have shown that, while the
center-of-mass energy of the colliding particles may diverge as
claimed in \cite{Ban1}, the maximal energy which is radiated to
infinity by an escaping particle (after the collision) is only one
order of magnitude larger than the initial energy of the colliding
particles (this energy gain reflects the extraction of rotational
energy from the spinning Kerr black hole).
\bibitem{Noteexm} As shown in \cite{Ban1,Ban2},
assuming the black hole to be a maximally-rotating (extremal) one,
allows one to maximize the center-of-mass energy of the colliding
particles.
\bibitem{Noteunit} We use natural units in which $G=c=\hbar=1$.
\bibitem{Notecf} It is worth mentioning that the first particle is
characterized by the critical angular momentum (\ref{Eq4}).
\bibitem{Notemq} It was shown in \cite{Ban4} that a particle infalling from spatial infinity
can reach the black-hole horizon provided $0\leq Q\leq 2\mu^2$.
\bibitem{Noteop} That is, the collision with the largest possible center-of-mass
energy.
\bibitem{Notemx} That is, the two colliding particles cannot reach
the near-horizon region $x_{\text{c}}<x^{\text{min}}_{\text{c}}$
without being engulfed by the new (and larger) horizon (\ref{Eq12})
which is formed according to the hoop conjecture \cite{Thorne} [and
also according to its weaker version (\ref{Eq10})].
\bibitem{Noterf1} Note that, due to the finite proper masses of the infalling
particles, the dimensionless critical angular momentum (\ref{Eq4})
receives a small correction term of order $O(\mu/M)$. Hence, one
should replace $l_1=2\to l_1=2[1+O(\mu/M)]$ in Eq. (\ref{Eq12}).
This, in turn, implies that the r.h.s of Eq. (\ref{Eq14}) should be
multiplied by the correction factor $1+O(\mu/M)$.
\bibitem{Noterf4} It is worth emphasizing that the expression (\ref{Eq16})
for the maximally allowed center-of-mass energy of the colliding
particles can be approached (from below) arbitrarily close when the
radial location of the collision approaches (from above) arbitrarily
close to $x^{\text{min}}_{\text{c}}$ [see Eq. (\ref{Eq13})].
\bibitem{Noterf3} It is worth emphasizing that the expressions (\ref{Eq14}) and (\ref{Eq16})
for the maximally allowed center-of-mass energy of the colliding
particles are valid for generic trajectories, whereas the value
(\ref{Eq20}) for $\gamma_{\text{max}}$ is obtained for a collision
at the equatorial plane \cite{Ban4}.
\bibitem{Noterf2} It is worth emphasizing again that, according to the hoop conjecture,
it is the presence of the particles themselves within a sufficiently
small radius that triggers the formation of a new horizon.
\bibitem{Noterf5} It is important to emphasize that Berti et. al. \cite{Beretal}
have shown that the gravitational radiation emitted by the infalling
particles is of order $E_{\text{rad}} \sim (\mu^2/M)\cdot
\ln(2-l_1)$. Since $2-l_1=O(\mu/M)$ (see \cite{Noterf1}), one finds
$E_{\text{rad}} \sim -(\mu^2/M)\cdot\ln(\mu/M)$, which implies that
the radiated gravitational energy can be neglected (that is,
$E_{\text{rad}}\ll\mu$) in the regime $-(\mu/M)\cdot\ln(\mu/M)\ll1$.
\bibitem{Noteweak} It is important to emphasize that, had we used the
original hoop conjecture \cite{Thorne} instead of its weaker version
(\ref{Eq10}), we would have found that the center-of-mass energy of
the colliding particles is bounded from above by a bound which is
{\it stronger} than our bound (\ref{Eq21}).
\end{thebibliography}
\end{document}